\documentclass[usenatbib]{mn2e}
\usepackage{astro_bib_macro}
\usepackage{txfonts}
\bibpunct{(}{)}{;}{a}{}{,}

\usepackage{epsf}

\title[A quick pseudo 3D photoionization code]{Modelling of aspherical nebulae. I. A quick pseudo-3D photoionization code}

\author[C. Morisset et al.]{C. Morisset$^{1}$\thanks{Morisset@Astroscu.UNAM.mx},
        G. Stasi\'nska$^{2}$,
        and M. Pe\~na $^{1}$ \\
        $^1$ Instituto de Astronom\'{\i}a, Universidad Nacional
           Aut\'onoma de M\'exico; Apdo. postal 70--264; Ciudad Universitaria;
           M\'exico D.F. 04510; M\'exico. \\ 
        $^2$ LUTH, Observatoire de Meudon, 5 place J. Janssen,
           F-92195 Meudon Cedex, France
    }

\newcommand{\ion}[2]{#1~\textsc{#2}}

\newcommand{\mysp}{}\def\mysp/{}
\newcommand{\mytimes}{}\def\mytimes/{}

\newcommand{\alloa}[3]{\ion{#1\mysp/}{#2}\ #3\AA}
\newcommand{\forba}[3]{[\ion{#1\mysp/}{#2}]\ #3\AA}

\newcommand{\dforba}[4]{[\ion{#1\mysp/}{#2}]\ #3,#4\AA}
\newcommand{\rforba}[4]{[\ion{#1\mysp/}{#2}]\ #3/#4\AA}
\newcommand{\uchii}{}\def\uchii/{UCHII}
\newcommand{\teff}{}\def\teff/{$\mathrm{T}_{\mathrm{eff}}$}
\newcommand{\kms}{}\def\kms/{km.s$^{-1}$}

\newcommand{\nebu}{}\def\nebu/{NEBU\_3D}
\newcommand{\hbeta}{}\def\hbeta/{H$\beta$}

\begin{document}

\maketitle

\begin{abstract}

We describe a  pseudo-3D photoionization code, \nebu/ and its associated visualization tool, VIS\_NEB3D, which are able to easily and rapidly treat a wide variety of nebular geometries, by combining models obtained with a 1D photoionization code. 
The only requirement for the code to work is that the ionization source is unique and not extended. It is applicable as long as the diffuse ionizing radiation field is not dominant and strongly inhomogeneous. 
As examples of the capabilities of these new tools, we  consider two very different theoretical cases. One is that of a high excitation planetary nebula that has an ellipsoidal shape with two polar density knots. The other one is that of a blister HII region, for which we have also constructed a spherical model (the spherical impostor) which has exactly the same H$\beta$ surface brightness distribution as the blister model and the same ionizing star.

We present and comment line intensity maps corresponding to different viewing angles. We  also use the computed line intensities to derive physical properties of the model in the same way as an observer would do for a real object. For example, we  derive  the ``apparent'' value of N/O for the entire nebulae and along spectral slits of different orientations. For this we take the electron temperature and density derived from the \forba{N}{ii}{5755}/\forba{N}{ii}{6583} and \forba{O}{ii}{3726}/\forba{O}{ii}{3729} ratios, respectively, and we adopt the common recipe: N/O=N$^+$/O$^+$. Interestingly, we find that, in the case of our high excitation nebula, the derived N/O is within 10-20\% of the real value, even when the slit crosses the high density knots. On the other hand, for the blister HII region and its spherical impostor, we find that the apparent N/O is much smaller than the true one (about 0.68 and 0.5 of it, respectively).

These two examples warn against preconceived ideas when interpreting spectroscopic and imaging data of HII regions and planetary nebulae. The tools \nebu/ and VIS\_NEB3D, which will be made publicly available in the future, should facilitate the performance of numerical experiments, to yield a better understanding of the physics of aspherical ionized nebulae.

\end{abstract}

\begin{keywords}
 methods: numerical -- ISM: H II regions -- planetary nebulae -- ISM: Abundances
\end{keywords}

            \section{Introduction}
\label{sec:intro}

Most ionized nebulae have complex morphological structures. This is true not only for HII regions, which result from the ionization of molecular clouds by newly born massive stars
\citep{Kim03,RFRR90,ODell01,MHT03},
but also for planetary nebulae \citep[see e.g. ][]{MGSS96,GSCW99,APN3}, not to mention images obtained by the Hubble Space Telescope which show a large variety of complex structures.

So far, almost all self-consistent physical analysis of ionized nebulae have used 1D photoionization codes. This is due to the fact that most photoionization codes are in 1D. In many cases, with appropriate handling, these codes are actually able to infer important conclusions regarding the ionization structure, the heating or the elemental abundances of the nebulae, even if the true geometrical structure is more complex  \citep{Clegg87,StS99,LP01,PaperIII,JSP04}.

A few 3D photoionization codes now exist \citep{BDG95,GVB97,Erc03,WME04}. Quite naturally, the first 3D photoionization models applied to real nebulae have  emphasized their success in reproducing selected line ratios, rather than their failure to reproduce all the observational constraints at the same time \citep[e.g. ][]{MMGV00,MSG04,MSGGH05}. However, reproducing simultaneously all the observational constraints, and not only strong line ratios, is paramount to check our understanding of the physics of these objects. Failure to reproduce let it be one line ratio, e.g. \ion{He}{ii}/H$\beta$ or \rforba{O}{iii}{4363}{5007} may indicate that an important ingredient is missing e.g. in the description of the ionizing radiation field or in the energy budget. Some recent work is starting to address  issues related to the physics of ionized nebulae in which a 3D treatment is important \citep{EMBSL03,EBSLRW03,WME04,Erc04}.

There are however practical difficulties with 3D model fitting of real nebulae. One is the enormous parameter space to consider, especially in the case of complex morphologies. Another one is the fact that 3D photoionization models are very CPU-time and memory consuming: several hours are needed on clusters or supercomputers to make a model 
(even if the time and memory can be reduced when modeling simpler geometries). 
Also, the difficulty in visualising the results does not help.

It would therefore be useful to have at one's disposal a quick 3D photoionization code coupled with an efficient visualisation tool. This would allow one to explore more efficiently a complex parameter space, in the search of a satisfactory solution. This would also be of important tutorial value for experiments aimed at estimating the effects of the nebular geometry on the derived physical parameters (e. g. chemical composition, ionizing radiation field) or at educating oneself about how a 3D physical object may appear to an observer under various viewing angles.

In this paper, we present a pseudo-3D photoionization code (\nebu/), which models aspherical planetary nebulae in a few minutes. The code uses various runs of a 1D photoionization code and constructs from them a 3D photoionization model. 
The main limitation comes from the fact that the diffuse radiation is not treated exactly, but in the "outward-only" approximation as designed by \citet{Tarter67}, where the diffuse radiation is assumed to be emitted outwards in a solid angle of 2$\pi$.
Its associated visualisation tool, VISNEB\_3D, allows one to easily plot diagrams that are useful for understanding the physical conditions inside the models and diagrams that can be directly compared to observations. 
This tool can also be used to post-process the 3D grids produced by proper 3D photoionization codes.

The code and the visualization tool are described in Section 2. The next sections are devoted to two applications: Section 3 treats the case of an ellipsoidal planetary nebula with two symmetrical polar knots and comments on the interpretation of observations of such knots in real nebulae in terms of chemical composition and excitation. Section 4 considers the case of a simple blister HII region  and compares it to a spherical HII region having the same H$\beta$ surface brightness distribution.  Conclusions and perspectives are given in Section 5.

            \section{\nebu/, a pseudo-3D photoionization code}
\label{sec:nebu3d}

\nebu/ is a pseudo-3D photoionization code. It consists in a set of tools, developed under IDL (RSI), used to construct a 3D model of an ionized nebula from a set of runs obtained by the 1D photoionization code NEBU \citep{mp96,P02}. It can easily be adapted to any other 1D photoionization code.

\nebu/ is thus not a full 3D photoionization code like Mocassin \citep{Erc03} for example or the code by \citet{WME04}. The main limitation of \nebu/ is in the treatment of the diffuse field, which is not fully consistent. For example, \nebu/ should not be used to model the ionization of shadows behind knots, as 1D models cannot account in an accurate way for the ionization of the shadowed region by radiation produced by the surrounding material.
On the other hand, when the radiation field is not changing drastically from one direction to the other, the use of a set of 1D models does not lead to strong errors. The main advantage of \nebu/ is that it is faster by 3 to 4 orders of magnitude than a full 3D code when running on the same computer.
Note also that its predictions are more accurate than those of the 3D photoionization codes of \citet{BDG95} or \citet{GVB97}.
The procedure to make a model with \nebu/ is the following. We first define a cube in which we place the nebula. For each cell of the cube, we define the polar coordinates, according to the size of the nebula we are modeling.

We run NEBU various times, changing the nebular density distribution at each run according to an angular variation law (NEBU also allows one to vary the chemical composition at any point in the nebula if one wishes to explore the effects of abundance inhomogeneities). The set of outputs (such as the electron temperature, the emissivities of selected lines at each radial position, etc...), which is defined according to the goal of the model, is then read and interpolated on the coordinate cube.
In fact, if the nebula is axisymmetric, we only need to compute the interpolated values on 1/8th of the cube, reconstructing the whole cube by rotating the results.
With such a procedure, one can  easily construct models of ellipsoidal or bipolar nebulae. One can also treat champagne flows or even irregular nebular geometries provided that the photoionization source is unique and that the effects of non-radial ionization are not expected to be important.

The resulting emissivity cubes (one for each line of interest) can then be rotated and a projection on the sky plane is done to obtain surface brightness maps.

We checked that we can recover the whole set of the 108 models published by \citet{ZK98} for the \hbeta/ images of idealized planetary nebulae, if adopting the same distribution for the gas.

            \section{Model for an ellispoidal planetary nebula with two polar knots}
\label{sec:model:elli}
            \subsection{Construction of the model and method of analysis}
\label{sub:model:elli:model}

A large fraction of planetary nebulae have an ellipsoidal appearance on the sky. Some of these, e.g. NGC 7009, NGC 6826, NGC 3242, present symmetric knots or ansae. It has been proposed that in some cases the knots may be nitrogen-rich \citep{BPMTH94,HA95}, although more recent analyses \citep{AB97,GC03,PP04} are much less categorical on this point. This  requires deeper investigation, as the ionization correction factors may be different in the knots and in the surrounding material, due to different ionization conditions. The model presented in this section illustrates this issue.

The density distribution is described by two components: a prolate ellispoidal nebula and polar knots.
The ellipsoidal nebula is defined by an inner ellipsoidal cavity $r_{in}(\theta)$ with an equatorial over polar axes ratio $b/a$ of 0.5. The density at the inner surface of the nebula is proportional to the inverse square of the distance to the ionizing star. Inside the nebula, the density decreases following a gaussian law.  The polar knots are defined by adding a gaussian  overdensity above the nebular background.

For a given polar angle $\theta$ between 0 and $\pi/2$ 
($\theta$ being 0 at the equator)
the density distribution at the distance $r$ from the ionizing source is then given by:

\begin{eqnarray}
    n_{\rm H}(\theta,r) =n_{\rm H}^{0} \mytimes/ \left( \frac{r_\mathrm{in}(0)}{r_\mathrm{in}(\theta)}\right)^{2}
        \mytimes/ \left(C^\mathrm{ellips}(\theta,r) + C^\mathrm{knot}(\theta,r)\right)
\end{eqnarray}
where $C^\mathrm{ellips}(\theta,r)$ and $C^\mathrm{knot}(\theta,r)$ are dimensionless functions of the position describing the ellipsoidal and knot contributions to the density, defined as:

\begin{eqnarray}
    C^\mathrm{ellips}(\theta,r > r_\mathrm{in}(\theta))  & =&
        \exp - \left(\frac{\Delta r(\theta)}{ S_\mathrm{size} \mytimes/ \Delta r_{Str}(\theta)} \right)^{2}
\\
    C^\mathrm{ellips}(\theta,r \leq r_{in}(\theta))  &=&  0
\\
    C^\mathrm{knot}(\theta,r)  &=&  C^\mathrm{knot}_\mathrm{max} \mytimes/
    \exp - \left(\frac{\Delta r(\theta) - K_\mathrm{pos} \mytimes/ \Delta r_{Str}(\theta)}{K_\mathrm{r-size} \mytimes/ \Delta r_{Str}(\theta)} \right)^{2}  \\
    && \mytimes/ \exp - \left(\frac{\theta - \pi/2}{K_\mathrm{\theta - size} /(\pi/2)}\right)^{2} \nonumber
\end{eqnarray}
using:
\begin{eqnarray}
    r_{in}(\theta) & = & r_\mathrm{in}(0) \mytimes/ \frac{a \mytimes/ b}
        {\sqrt{\left((b \mytimes/ \sin(\theta))^{2}+(a \mytimes/ \cos(\theta))^{2}\right)}}
\\
    \Delta r(\theta) & = & r - r_{in}(\theta)
\end{eqnarray}

The relative intensity of the knot contribution is defined using $C^\mathrm{knot}_\mathrm{max}$.
The size scales of the nebula and the knots are defined relatively to the thickness of an ionized shell $\Delta r_{Str}(\theta)$\footnote{In the case of a constant density shell around a cavity, we have: $\Delta r_{Str} = r_{in}((\frac{3 Q_0}{4\pi \alpha_B n_{\rm H}^{2} r_{in}^{3} }+1)^{1/3} - 1) $} corresponding to a constant density $n_{\rm H}(\theta,r_{in}(\theta))$ around an empty cavity of size $r_{in}(\theta)$. Dimensionless coefficients are applied to these $\Delta r_{Str}(\theta)$, namely $S_\mathrm{size}$ which is used for the gaussian decrease of density in the ellipsoidal contribution, $K_\mathrm{pos}$ used for the position of the knot on the polar axes within the nebula and $K_\mathrm{r-size}$ used for the radial size of the knot. Finally, $K_\mathrm{\theta - size}$ is the angular size of the knot, normalized to $\pi/2$.  The reason for using such a density law is that a change in the physical parameters ($n_{\rm H}^{0}$, $r_\mathrm{in}(0)$ and $L_*$) whilst keeping the dimensionless coefficients constant allow us to change the ionization parameter keeping the global shape of the nebula.

The model presented in this section is obtained using the following values for the physical parameters: $n_{\rm H}^{0} = 2\, 10^{3}$~cm$^{-3}$, $r_\mathrm{in}(0)=10^{17}$~cm, and the following values for the dimensionless parameters: $a=2, b = 1$, $C^\mathrm{knot}_\mathrm{max}=5.0$, $S_\mathrm{size}=1.7$, $K_\mathrm{pos}=0.3$, $K_\mathrm{r-size}=0.1$, and $K_\mathrm{\theta - size}=0.078$. 

We have adopted the following chemical composition for the entire nebula, including the knots: 
H:He:C:N:O:Ne:S:Ar=
1:0.1:3 10$^{-4}$:1 10$^{-4}$:6 10$^{-4}$:1.5 10$^{-4}$:1.5 10$^{-5}$:3 10$^{-6}$, by number.

The characteristics of the ionizing star are  $T_{eff}=80$~kK and $L_*=10^{36}$~erg/s.

We have run 40 models with NEBU to cover the domain $[0,\pi/2]$ for $\theta$. The number of steps for the NEBU runs is between 60 and 85, depending on the density distribution associated with each angle.

The results of the NEBU runs are then interpolated in the coordinate cube. The cube in which the interpolation id done is composed of 200$^3$ cells. For the nebula considered here, this correspond to a size for each cell of 2.28 10$^{15}$~cm. The variables that are interpolated are: the electron temperature and density, the hydrogen density, the ionic abundances of N$^+$ and O$^+$, and the emissivities of the line of interest, namely H$\beta$, \alloa{He}{ii}{4686}, \forba{N}{ii}{5755}, \forba{N}{ii}{6583}, \forba{O}{ii}{3726}, \forba{O}{ii}{3729} and \forba{O}{iii}{5007}.

\begin{figure}
\epsfxsize=10.0cm \hspace{-1cm} \epsfbox{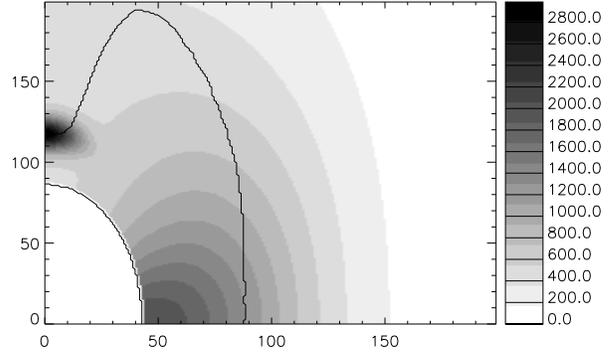}
\caption{Cut in the hydrogen density distribution for the model presented in Sect.~\ref{sub:model:elli:model}. The solid line indicates the recombination front of the nebula. The grey scale is in cm$^{-3}$.\label{fig:imdens}}
\end{figure}

\begin{figure}
\epsfxsize=9.0cm \epsfbox{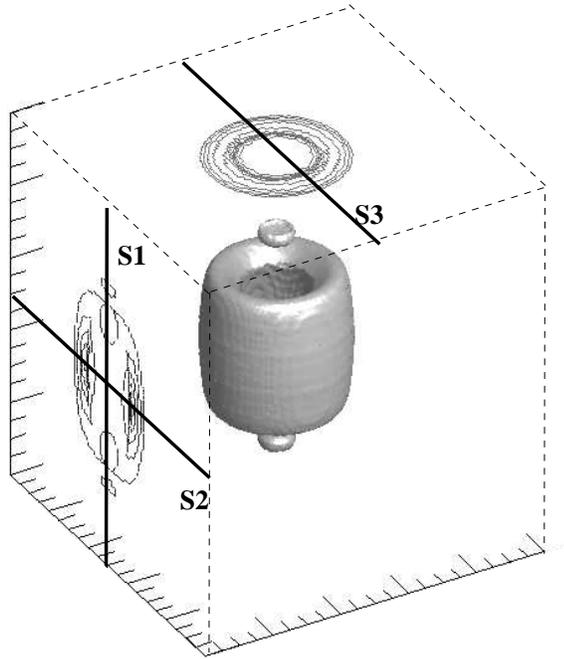}
\caption{3D representation of the nebula. An isodensity surface is drawn, showing the equatorial density enhancement and the two polar knots. On the faces we have represented the H$\beta$ surface brigthness contours for 2 orientations of the nebula: on the left side for the nebula seen with the polar axis paralle to the sky, on the top for the nebula seen pole-on. The slits used to determine parameters from emission line ratios are indicated.\label{fig:im1}}
\end{figure}

Figure~\ref{fig:imdens} is a 2D map of the  hydrogen density distribution of our model in a plane containing the polar axis. The star is at coordinates (0,0), and the recombination front is also plotted. Figure~\ref{fig:im1} is a 3D isodensity representation of the model. The chosen density level allows one to clearly distinguish  the main body of the nebula and the polar knots.

After having computed the model, we generate line intensity maps corresponding to ``pole-on'' and ``face-on'' viewing angles, by summing up the corresponding emissions on each line of sight. An example of such ``pole-on'' and ``face-on'' contour maps is shown in Fig.~\ref{fig:im1}.
Line ratio maps are then obtained by performing a pixel by pixel division of the two line intensity maps.

We then consider the model as if it were a real nebula, and apply to it the standard abundance analysis using  the same atomic data as in NEBU. 
For each pixel of the \rforba{N}{ii}{5755}{6583} map we compute the ``apparent'' electron temperature and for each pixel of the \rforba{O}{ii}{3726}{29} we compute the ``apparent'' electron density. These values are used to compute for each line of sight the ``apparent''  N$^+$/O$^+$ ratio from the \forba{N}{ii}{6583}/ \dforba{O}{ii}{3726}{29} maps, which is identified to  N/O, following the common practice. 

            \subsection{The results}
\label{sub:model:elli:res}

            \subsubsection{Synthetic nebular images }

\begin{figure}
\epsfxsize=12.0cm \hspace{-2cm} \epsfbox{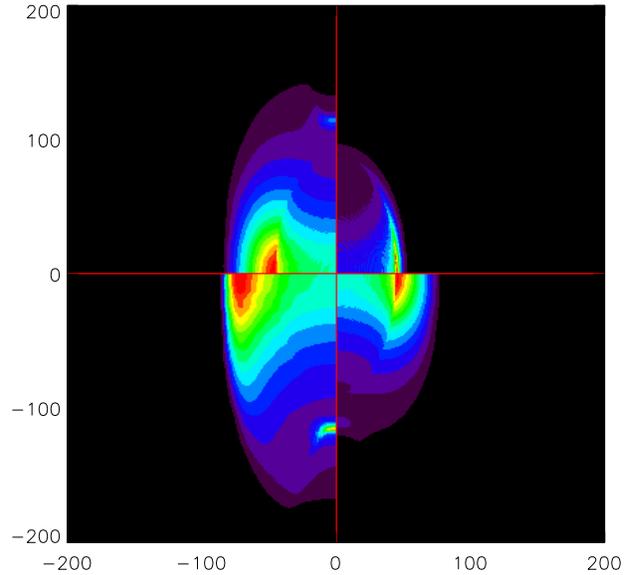}
\caption{Synthetic intensity maps of the ellipsoidal nebula when seen perpendicular to the polar axis. Each quadrant corresponds to an emission line: Upper left: H$\beta$, upper right: \alloa{He}{ii}{4686}, lower left: \dforba{O}{ii}{3726}{29}, and lower right: \forba{O}{iii}{5007}. The intensity maps are normalized so that the maximum for each line is the same.\label{fig:res1}}
\end{figure}

Fig.~\ref{fig:res1} shows  intensity maps of the model nebula seen in a direction perpendicular to the polar axis. Four emission lines are represented: H$\beta$, \alloa{He}{ii}{4686}, \dforba{O}{ii}{3726}{29}, and \forba{O}{iii}{5007}. Such a  presentation allows one to distinguish immediately various characteristics of the model images. For example, it is clearly seen that the  H$\beta$ and \dforba{O}{ii}{3726}{29} images are more extended than the  \alloa{He}{ii}{4686} one, and that the extension of the  \forba{O}{iii}{5007} is intermediate. One also sees that the emission is stronger along the equator. This is due to the density being smaller at high polar angle. 
The inner ``egg-shape'' seen in H$\beta$ and \forba{O}{iii}{5007} results from the elipsoidal inner empty cavity.
The polar knots are well visible in the H$\beta$ map and even more conspicuous in the \dforba{O}{ii}{3726}{29} map (the same applies for  the \forba{N}{ii}{6583} line map, not shown here), due to enhanced recombination in the knots. 
Even if the model presented in this section does not attempt to reproduce a ``real'' nebula, such emission enhancement of the low ionization species can be seen in real nebulae showing ANSAE such as NGC~7009 \citep{GC03}.

            \subsubsection{Local properties}

\begin{figure}
\epsfxsize=9.0cm \epsfbox{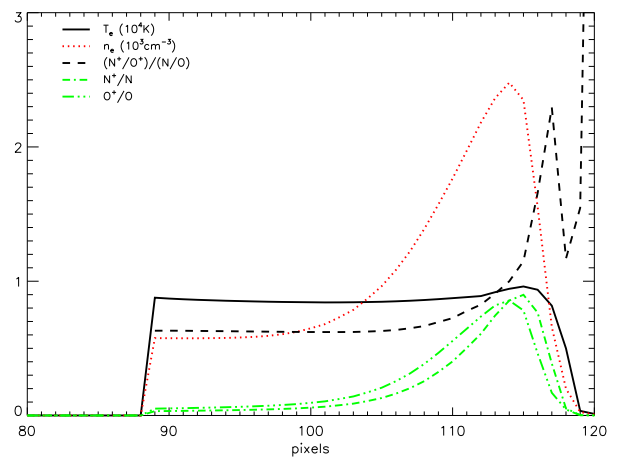}
\caption{Variation of some \emph{local} properties in the 3D grid of the model, along the large axis of the nebula (i.e. crossing the knots). Solid line: $T_{\rm e}$ (10$^4$K), dotted: $n_{\rm H}$ (10$^3$~cm$^{-3}$), dashed: (N$^+$/O$^+$)/(N/O), dashed-dot: N$^+$/N, and dot-dot-dashed: O$^+$/O. Note that in this figure the pixels range from 80 to 120. \label{fig:res2}}
\end{figure}

\begin{figure}
\epsfxsize=9.0cm \epsfbox{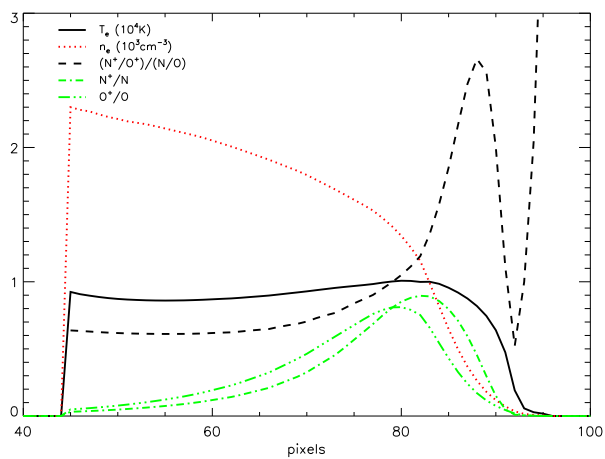}
\caption{Same as Fig.~\ref{fig:res2} but along a small axis of the nebula. Note that in this figure the pixels range from 40 to 90. \label{fig:res3}}
\end{figure}

Figs.~\ref{fig:res2} and~\ref{fig:res3} show the variation of some local properties in the model grid ($T_{\rm e}$, $n_{\rm e}$, (N$^+$/O$^+$)/(N/O), N$^+$/N, and  O$^+$/O) along the polar and equatorial axes respectively. We see that the maximum density of the knot is larger by a factor about 3 than the inner density along the polar axis (Fig.~\ref{fig:res2}) and only sligthly larger than the inner density on the equatorial plane (compare the dotted line in Figs.~\ref{fig:res2} and \ref{fig:res3}).  Note that the zone of maximum density in the knots lies out of the ionized region. The electron temperature does not show significant structure apart from the natural drop in the recombination zone (see the solid line in  Figs.~\ref{fig:res2} and \ref{fig:res3}). The ionization fractions  N$^+$/N (dashed-dot line) and  O$^+$/O (dot-dot-dashed line) increase away from the central star, as naturally expected, with the increase being more important in the zone of the dense knot. The (N$^+$/O$^+$)/(N/O) ratio (dashed line), which is generally assumed equal to one in abundance analysis, is actually never strictly equal to one in our model. It is of the order of 0.6 -- 0.7 in the high ionization zone and increases to values of about 2.5 on the back side of the knot as well as in the ionization front. In the latter case, the electron temperature is already low, so that this increase will have no effect on observed line ratios. The strong maximum at the back side of the knot is probably exagerated in our model, since the diffuse ionizing radiation is not treated properly. 

            \subsubsection{Line of sight intensities}

\begin{figure}
\epsfxsize=9.0cm \epsfbox{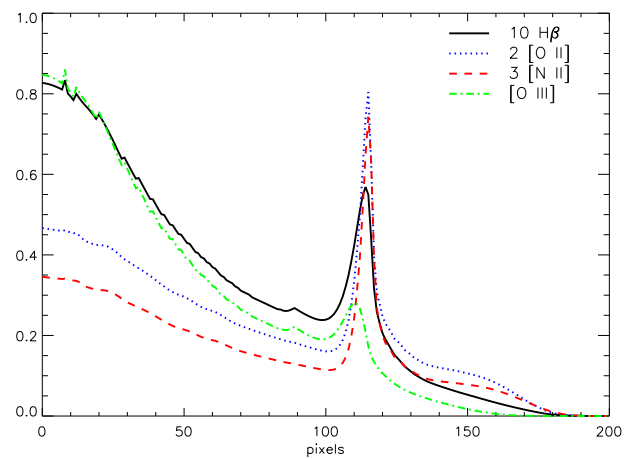}
\caption{Variation of the intensity of some lines along a narrow slit oriented from the ionizing star to the polar knot (slit S1 in Fig.~\ref{fig:im1}). Solid: 10$ \times$ H$\beta$, dotted: 3$ \times$ \dforba{O}{ii}{3726}{29}, dashed: 3$ \times$ \forba{N}{ii}{6583} and dot-dashed: \forba{O}{iii}{5007}. The knot is in the vicinity of pixel 110\label{fig:res4}}
\end{figure}

\begin{figure}
\epsfxsize=9.0cm \epsfbox{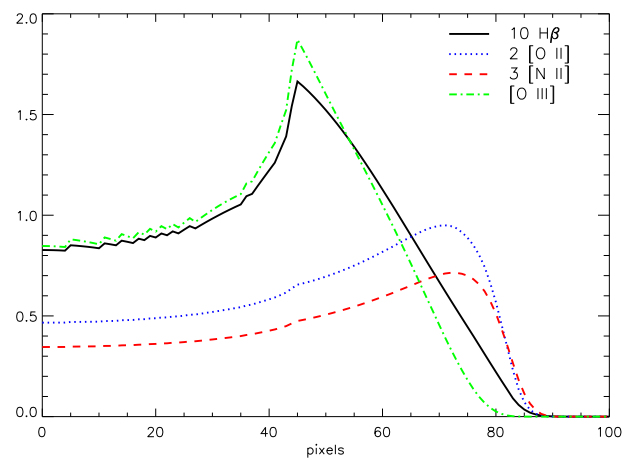}
\caption{Same as Fig.~\ref{fig:res4}, but through a narrow slit in the equatorial plane (slit S2 in Fig.~\ref{fig:im1}). \label{fig:res5}}
\end{figure}

\begin{figure}
\epsfxsize=9.0cm \epsfbox{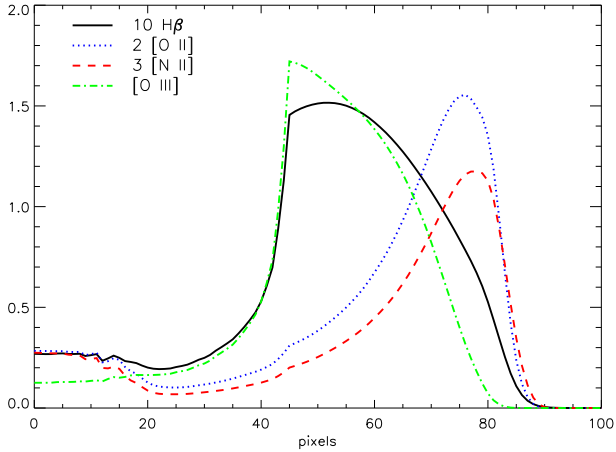}
\caption{Same as Fig.~\ref{fig:res4}, but for the nebula seen ``pole-on'' through a narrow slit passing through the center of the knots.  (slit S3 in Fig.~\ref{fig:im1}). The knot is centered at pixel 0. \label{fig:res6}}
\end{figure}

Figs.~\ref{fig:res4} to~\ref{fig:res6} show how the above local properties translate into the intensities of emission lines as would be observed with a narrow slit. Fig.~\ref{fig:res4} corresponds to the case where the slit is on the large axis and crosses the knots (slit S1 in Fig.~\ref{fig:im1}). Fig.~\ref{fig:res5} corresponds to the case where the slit is along the small axis (slit S2 in Fig.~\ref{fig:im1}). Fig.~\ref{fig:res6} corresponds to the case where the nebula is seen pole-on with a slit passing through the center  of the knots (slit S3 in Fig.~\ref{fig:im1}).
 The intensity axes are in arbitrary units, but the relative line intensities  are meaningful (applying the coefficients given in the caption). Fig.~\ref{fig:res4} shows that all the line of sight intensities present an enhancement at the position of the knot. However, the enhancement is moderate in the case of \forba{O}{iii}{5007} (dot-dashed line) due to the opposite effects of density enhancement and recombination of O$^{++}$ in the knot. In the case of H$\beta$ (solid line), the intensity is roughly proportional to the integral of the squared density along the line of sight, so the enhancement is larger than for \forba{O}{iii}{5007} (with respect to the nearby lines of sight, the enhancement is of about a factor 3 while it is only of 50\% for \forba{O}{iii}{5007}).  The intensities of \forba{N}{ii}{6583} (dashed line) and \dforba{O}{ii}{3726}{29} (dotted line) are more strongly enhanced (by about a factor 4--5) because the N$^{+}$/N and O$^{+}$/O ratios are larger in the knot than in the surrounding material. The inner cavity and the equatorial density enhancement are clearly seen in spectra along the small axis and when the nebula is seen pole-on (see Figs.~\ref{fig:res5} and~\ref{fig:res6}). On the other hand, they are not seen when the slit is along the large axis (Fig.~\ref{fig:res4}), because the lack of emission in the cavity is compensated by the stronger emissivity in the equatorial plane where the gas is overall denser than on the polar axis. Note that in our model, when the nebula is seen pole-on (Fig.~\ref{fig:res6}), the bump corresponding to the knot is barely seen and the emission is dominated by photons arising from dense regions from the body of the nebula for which the emission measure is large in such a configuration.

            \subsubsection{Apparent properties }

\begin{figure}
\epsfxsize=9.0cm \epsfbox{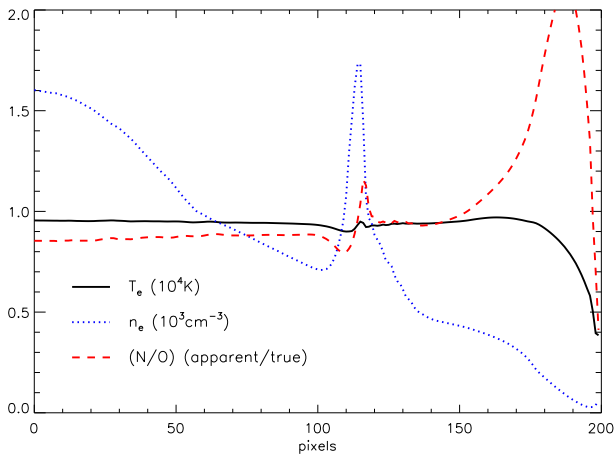}
\caption{Apparent variation of some properties derived from the emission line ratios along a narrow slit oriented along the polar axis and crossing the knot (slit S1 in Fig.~\ref{fig:im1}). Solid line: $T_{\rm e}$ (10$^4$K), dotted: $n_{\rm e}$ (10$^3$~cm$^{-3}$), dashed: N/O normalized to the ``true'' value of 1/6.\label{fig:res7}}
\end{figure}

\begin{figure}
\epsfxsize=9.0cm \epsfbox{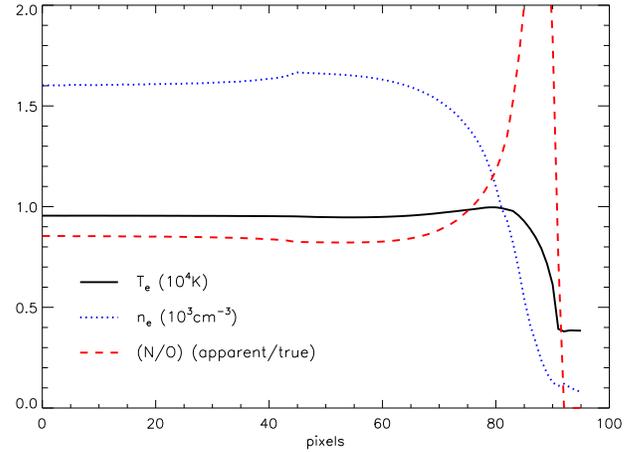}
\caption{Same as Fig.~\ref{fig:res7}, but for a a narrow slit oriented along the small axis (slit S2 in Fig.~\ref{fig:im1}).\label{fig:res8}}
\end{figure}

\begin{figure}
\epsfxsize=9.0cm \epsfbox{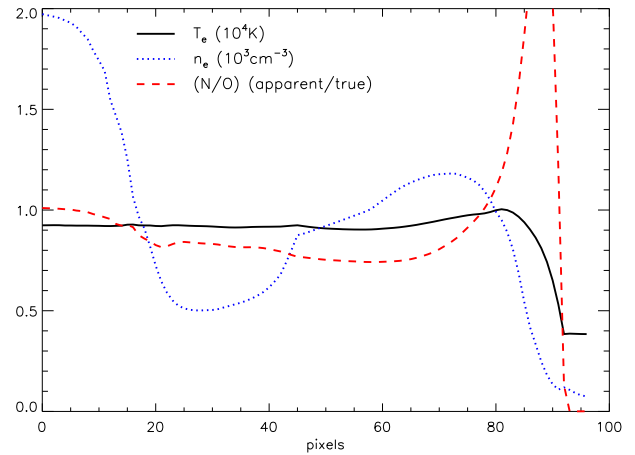}
\caption{Same as Fig.~\ref{fig:res7}, but for a ``pole-on'' nebula (slit S3 in Fig.~\ref{fig:im1}).\label{fig:res8b}}
\end{figure}

Figs.~\ref{fig:res7} to~\ref{fig:res8b} show the values of  $T_{\rm e}$ (solid line), N$_{\mathrm e}$ (dotted line) and N/O (dashed line) determined using the  line of sight intensities of \forba{N}{ii}{5755}, \forba{N}{ii}{6583}, \forba{O}{ii}{3726} and \forba{O}{ii}{3729} lines in Figs. ~\ref{fig:res4} to~\ref{fig:res6}, respectively. The procedure to derive these quantities has been described in Sect.~\ref{sub:model:elli:model} and corresponds to what an observer would do using the line intensities from long slit spectra of a real object. We call these quantities ``apparent'' properties of the model.

As seen in Fig.~\ref{fig:res7}, in the case of a slit along the polar axis, the apparent electron density decreases steadily from the center until the position of the knot is reached where a sharp peak is seen. On the other hand, when the nebula is observed through a slit aligned on the small axis (Fig.~\ref{fig:res8}), the apparent density is constant over most of the nebular apparent surface except the outermost zone where it decreases abruptly. The presence of the central cavity, of course, cannot be guessed from such line ratios, and can only be inferred from the radial distribution of line fluxes such as presented in Figs.~\ref{fig:res4} to~\ref{fig:res6}. The fact that the apparent density is uniform if the nebula is seen along a small axis is simply due to the fact that most of the \dforba{O}{ii}{3726}{29} emission comes from a thin external zone. On the other hand, along the polar axis, the apparent density in the central parts is strongly affected  by the dense equatorial zone, and therefore the apparent density decreases outwards.

The apparent N/O value (obtained assuming N/O=N$^+$/O$^+$) is never identical to the true value. As seen in Figs.~\ref{fig:res7} and~\ref{fig:res8}, the ratio  (apparent N/O) / (true N/O), i.e. the dashed lines is 0.85 in the whole nebula, except close to the knot, where it drops to 0.78 before increasing up to a value of 1.2 at the far side of the knot. This means that from the observations of the emission lines shown in Fig.~\ref{fig:res4}, one would infer a modest variation (up to 40\% in this model) of the N/O ratio across the nebula. Taking into account the slit width and the seeing, an observer would probably not distinguish any abundance variation in such a case.  Notice that the apparent increase of N/O at the edge of the nebula to values up to 2.5 times the true value (pixel numbers greater than 160 and 80 in Figs.~\ref{fig:res7} and~\ref{fig:res8} resp.) are not relevant, as the emission of all the lines is very low at this position in the nebula (see Figs.~\ref{fig:res4} and~\ref{fig:res5} resp.), and such an increase could not be seen.

\begin{figure}
\epsfxsize=9.0cm \epsfbox{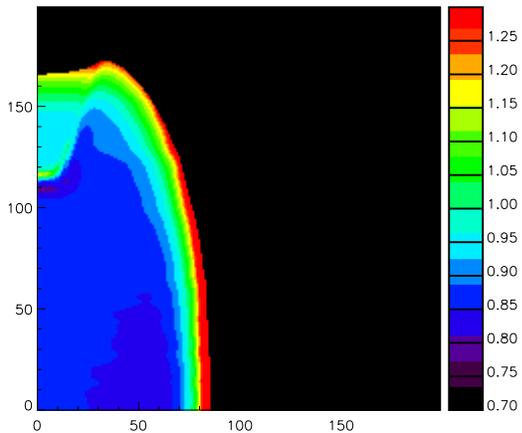}
\caption{Image of (apparent N/O) / (true N/O) obtained from the emission line maps of Fig.1.\label{fig:res9}}
\end{figure}

Fig.~\ref{fig:res9} shows the map of the bias in N/O for the case of a nebula whose polar axis is in the plane of the sky.  The pixels where the intensity of \dforba{O}{ii}{3726}{29} is lower than 0.05 times its maximum value are set to black, to avoid misinterpretation of the increase of N/O at the edge of the nebula, where no emission arises. Similarly to what is seen in Fig.~\ref{fig:res8}, we see the apparent increase of N/O just behind the knot.

\begin{figure}
\epsfxsize=9.0cm \epsfbox{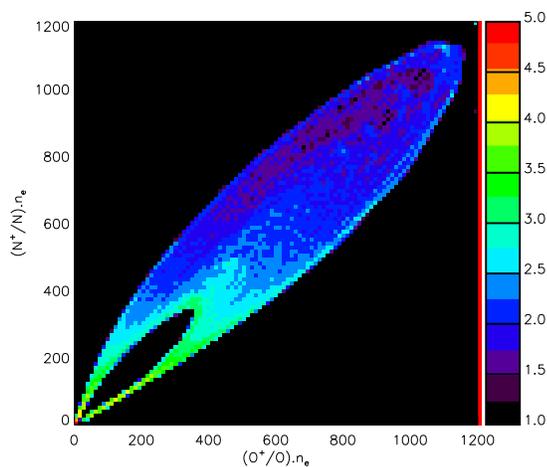}
\caption{2D histogramm of the number of cells with a given value of (N$^+$/N) $n_{\rm e}$ and (O$^+$/O) $n_{\rm e}$, for all the cells in the nebula where O$^{+}$/O $>$ 0.02. The vertical scale on the right gives the correspondance between the grey level and the log of the number of cells. \label{fig:res10}}
\end{figure}

Fig.~\ref{fig:res10} 2D histogramm of the number of cells with a given value of $n_{\rm e}$~$n$(N$^+$) and $n_{\rm e}$~$n$(O$^+$), for all the cells in the nebula where O$^{+}$/O $>$ 0.02. The vertical scale on the right gives the correspondance between the grey level and the log of the number of cells. 
The parts of the nebula emitting strongly in \dforba{O}{ii}{3726}{29} and \forba{N}{ii}{6583} are far from the (0,0) origin, the parts emitting weakly in these lines are close to it. 
If the assumption that N/O=N$^+$/O$^+$ were true, then all the points would lie on the y=x axis. That is clearly not the case, the majority of points are either under or above the first bisector. 

Note that the N/O derived for the entire nebula from an integrated spectrum is 0.93 in this case, which happens to be very close to the true value, due to counterbalancing effects in the various zones.
This example shows  that, even if N$^+$ and O$^+$ are present in small proportion in a high excitation nebula, the overall N/O abundance ratio derived by classical empirical methods may be  reasonably accurate. 

Whether our results on the apparent N/O value are valid in general (or at least for the majority of planetary nebulae) would require a scan of the entire parameter space.

            \section{Models for a blister HII region and its spherical impostor}
\label{sec:model:blister}

The idea that HII regions may be described by a champagne flow model has been widespread since the paper by \citet{Z73} who showed that the Orion nebula is not a sphere but has the structure of a blister, and the theoretical paper by \citet{TT79}, who showed how such a structure is likely to arise. However, very little photoionization work for such geometries has been performed. Among the few exceptions are the work of \citet{BGM03} on the Orion nebula, and that of \citet{PaperIII} on G29.96-0.02 to date.

As a matter of fact, most modelling is still done either in plane parallel approximation, or in spherical symmetry. Here, we construct a photoionization model for a blister HII region ionized by a single star. The H$\beta$ image of such a model when seen face on is circular, and the object could be mistaken for a sphere.

In this section, we discuss the properties of a blister model and compare them with those of a spherical model that has the same ionizing star, the same chemical composition and the same total H$\beta$ luminosity and surface brightness distribution. We call this second model the spherical impostor.

            \subsection{The models}
\label{sub:model:blister:mod}

The blister model presented here is a plane parallel slab, with density increasing exponentially with the distance to the ionizing star, as found by  \citet{WD95} for the Orion nebula: $n_{\rm H} (z) = n_{\rm H}^{0} \times \exp((z-z_b)/L)$, where $z$ is the normal distance to the star and $z_b,L$ are parameters taken to be equal to 2 10$^{18}$ and 5 10$^{17}$~cm respectively. The value of n$_{\rm H}^{0}$ in our model is 3 10$^{4}$~cm$^{-3}$. The star is assumed to radiate as a blackbody with an effective temperature  of T$_{\mathrm eff}= 37000$~K  and a luminosity of 3.5 10$^{39}$erg/s. The chemical composition is solar.
The gas density distribution law of the spherical impostor has been obtained by trial and error and is the following: 
$ n_{\rm H}(r) = n_{\rm H}^{0} \exp(-(r/C)^{2}) \times B^{2}/(r^{2}+D^{2})$, where $n_{\rm H}^{0}=2.6 10^{3}$~cm$^{-3}$, B=1.5 10$^{18}$~cm, C=1.2 10$^{18}$~cm, and D=6 10$^{17}$~cm. An empty inner cavity around the ionizing star of radius 10$^{17}$~cm is assumed.

            \subsubsection{Spatial distribution of the line intensities }

\begin{figure*}
\epsfxsize=18.0cm \epsfbox{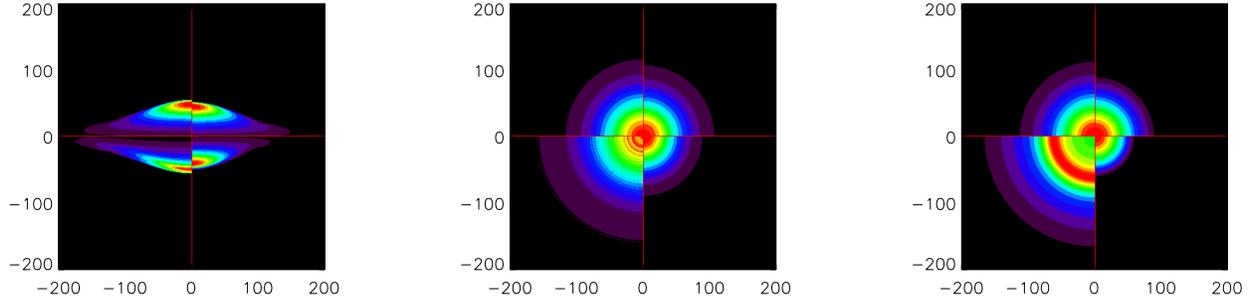}
\caption{Synthetic H$\beta$, \alloa{He}{i}{6678}, \dforba{O}{ii}{3726}{29} and \forba{O}{iii}{5007} surface brightness maps from upper left to lower right. Left panel: the blister model seen tangentially (in this case the maps for \dforba{O}{ii}{3726}{29} and \forba{O}{iii}{5007} are upside down); central panel: the blister model seen face on; right panel: the spherical impostor model.\label{fig:res4imbli}}
\end{figure*}

Figure~\ref{fig:res4imbli} presents line intensity maps of the models, in a way similar to Fig.~\ref{fig:res1}, but with   \alloa{He}{ii}{4686} replaced by \alloa{He}{i}{6678}, since the stellar temperature is only of $37000$~K. The left panel corresponds to the blister model, seen tangentially to the slab. In this panel the \dforba{O}{ii}{3726}{29} and \forba{O}{iii}{5007} maps are upside down, for a better comparison. Note that the blister model is not a plane parallel model, since the radiation comes from a star whose distance to the slab is not infinite. We see that, for each line, the surface brightness increases towards the ionization front. This is due to the adopted density law, which increases with distance to the star. We also see that the surface brightness close to the ionization front is larger in the direction perpendicular to the slab, and decreases with increasing polar angle. This is due to the dilution of the radiation which is more important at higher polar angles, so that the ionization front occurs in a less dense zone. As expected, the extensions of the  \forba{O}{iii}{5007} and \alloa{He}{i}{6678} emitting zones are smaller than those of the \dforba{O}{ii}{3726}{29} and H$\beta$ emitting zones. This is better seen in the maps corresponding to the face-on blister, which are presented in the central panel of Fig.~\ref{fig:res4imbli}. The right panel of Fig.~\ref{fig:res4imbli} presents the line intensity maps of the spherical impostor model. By construction, the H$\beta$ map is identical to that of the face-on blister. 

\begin{figure*}
\epsfxsize=18.0cm \epsfbox{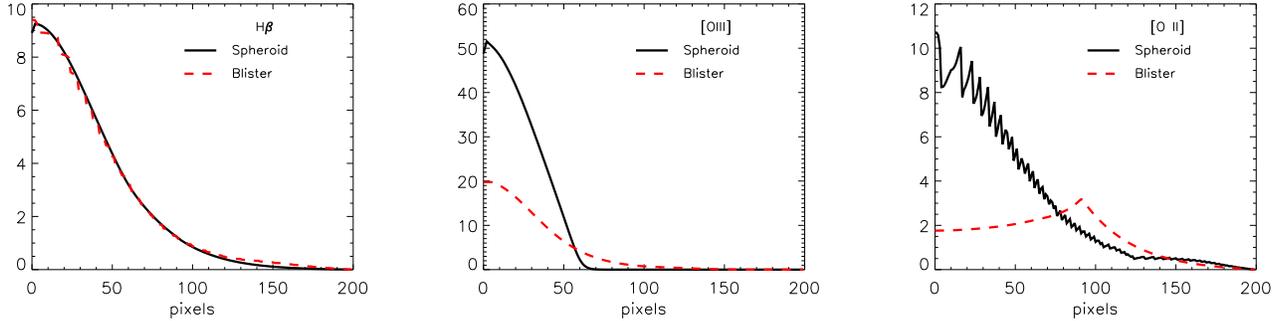}
\caption{Emission line intensity variations (from left ro right: H$\beta$, \forba{O}{iii}{5007}, \dforba{O}{ii}{3726}{29}). Solid line: spherical model, dashed line: blister model.\label{fig:sphbli1}}
\end{figure*}

The maps for the remaining lines are significantly different. They are better discussed by considering also Fig.~\ref{fig:sphbli1}, which compares the H$\beta$, \dforba{O}{ii}{3726}{29} and \forba{O}{iii}{5007} line intensity variations for the face-on blister (dotted lines) and for the spherical impostor (solid line). We seen that the excitation level of the spherical impostor is much higher than that of the blister. The reason is that, in the spherical impostor, the dense matter is close to the star and receives a strong radiation field, while in the blister, the dense matter is far from the star. In the spherical impostor, the \forba{O}{iii}{5007} intensity decreases rapidly away from the central line of sight, because the O$^{+}$ ionizing photons are quickly exhausted due to the dense inner region, while in the face-on blister, the variation of the ionizing radiation field penetrating the gas is much smoother. The \dforba{O}{ii}{3726}{29} map also shows conspicuous differences between the two cases\footnote{In the case of the blister model, the \dforba{O}{ii}{3726}{29} lines are not smooth for numerical reasons, the size of the region emitting these lines beeing of the order of a few cells (See Fig.~\ref{fig:res4imbli}). We checked that the values are close to the outputs of the original NEBU runs.}. In the spherical impostor, the curves related to the \dforba{O}{ii}{3726}{29} intensity first gently increases away from the central line of sight, due to the radial increase of the O$^{+}$ population convolved with the decrease of the gas emission measure on the line of sight (the integral of the square of the gas density). Farther away,  all the oxygen is in the O$^{+}$ form, and the \dforba{O}{ii}{3726}{29} intensity decreases. On the other hand, for the face-on blister, the \dforba{O}{ii}{3726}{29} intensity decreases steadily outwards. Indeed, the effect of the  emission measure decreasing with increasing polar angle dominates over the ionization effect. This result is of course strongly dependent on the gas density law taken for the slab. With a much weaker density increase, one might observe a behaviour qualitatively similar to the one shown by the solid line in the right panel of Fig.~\ref{fig:sphbli1}.

\begin{figure*}
\epsfxsize=18.0cm \epsfbox{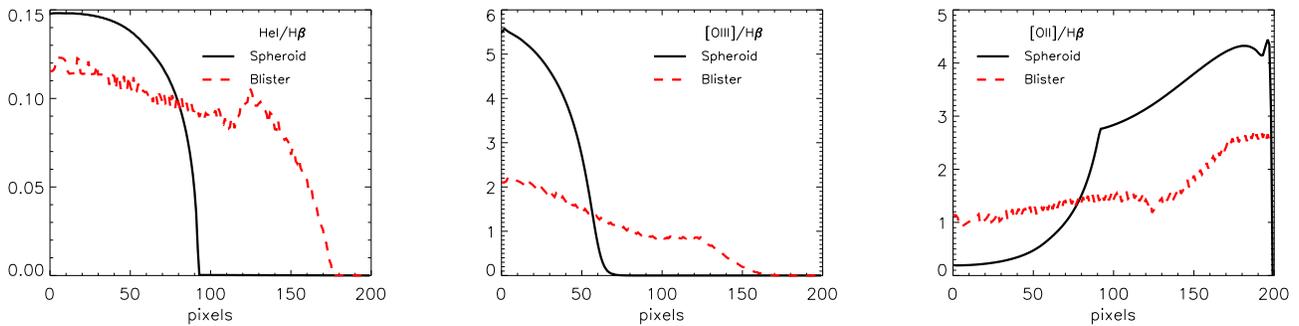}
\caption{Emission line intensity variations, normalized to H$\beta$ (from left ro right: \alloa{He}{i}{5876}, \forba{O}{iii}{5007}, \dforba{O}{ii}{3726}{29}). Solid line: spherical model, dashed line: blister model.\label{fig:sphbli2}}
\end{figure*}

Figure~\ref{fig:sphbli2} is similar to Fig.~\ref{fig:sphbli1}, but shows line ratios instead of line intensities: from left to right, \alloa{He}{i}{5876}/H$\beta$, \forba{O}{iii}{5007}/H$\beta$ and \dforba{O}{ii}{3726}{29}/H$\beta$. We see that  \forba{O}{iii}{5007}/H$\beta$ decreases abruptly for the spherical impostor, while it decreases very slowly for the blister. Similarly, \dforba{O}{ii}{3726}{29}/H$\beta$ increases abruptly for the spherical impostor, and slowly for the blister. Indeed, the ionization parameters of the sampled zones are much higher in the spherical impostor than in the face-on blister, resulting in a much higher central \forba{O}{iii}{5007}/H$\beta$ ratio and a sharper ionization stratification.  As for the \alloa{He}{i}{5876}/H$\beta$ ratio, it decreases outwards, as expected. In the spherical impostor, \alloa{He}{i}{5876}/H$\beta$ abruptly goes down to zero at about 90 pixels, while the decrease is much slower in the face-on blister, again because of the sharper ionization stratification in the spherical impostor. The total \alloa{He}{i}{5876}/H$\beta$ value is 0.029 for the spherical impostor , i.e. slightly larger than the value of 0.027 for the blister.

This example shows that, if one were to construct a photoionization model for a blister HII region seen face on, and assumed a spherical geometry, one would run into severe difficulties matching the observed line ratios as a function of projected distance to the star. 

            \subsubsection{Comparison of apparent properties }

\begin{figure*}
\epsfxsize=18.0cm \epsfbox{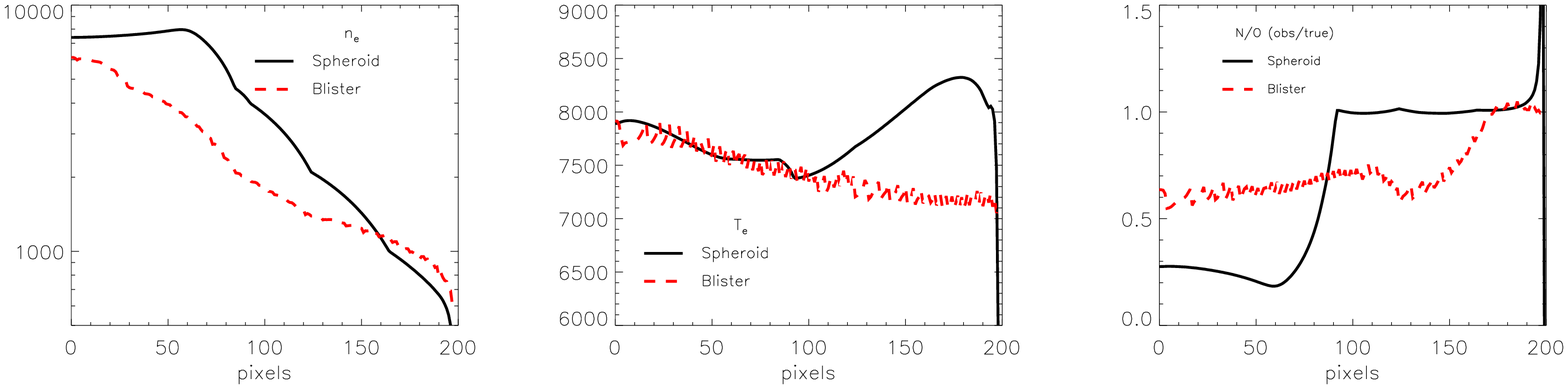}
\caption{Variation of some ``apparent'' parameters (from left ro right: $n_{\rm e}$, $T_{\rm e}$, and N/O), determined from the emission line ratios shown in Fig.~\ref{fig:sphbli1}. Solid line: spherical model, dashed line: blister model.\label{fig:sphbli3}}
\end{figure*}

\begin{figure}
\epsfxsize=9.0cm \epsfbox{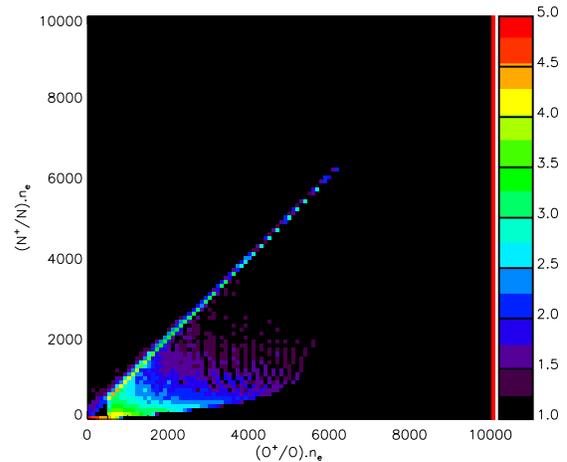}
\caption{Same as Fig.~\ref{fig:res10}, but for the blister model.\label{fig:bli:hist}}
\end{figure}

\begin{figure}
\epsfxsize=9.0cm \epsfbox{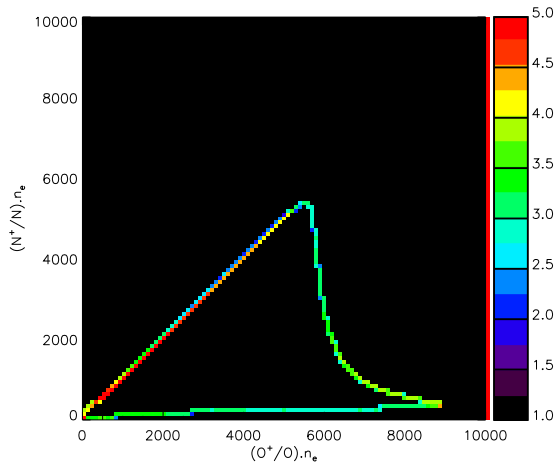}
\caption{Same as Fig.~\ref{fig:res10}, but for the spherical impostor model.\label{fig:sph:hist}}
\end{figure}

We now compare the apparent physical properties of the models ($T_{\rm e}$, $n_{\rm e}$, and N/O) derived from empirical methods in a way similar to what we presented in Sect.~\ref{sub:model:elli:model}. Figure~\ref{fig:sphbli3} shows the electron density derived from \rforba{O}{ii}{3726}{29} (left panel), the electron temperature derived from \rforba{N}{ii}{5755}{6583} (middle panel) and the empirically derived N/O ratio (right panel). In both models, we observe a general decrease of the apparent electron density away from the star. In the case of the blister, this  apparent density decrease is due to the fact that, as the polar angle increases, the regions that are sampled by the \dforba{O}{ii}{3726}{29} lines are of decreasing density, due to the increasing dilution of the ionizing radiation field. Note that the apparent density decrease we find for our model is very similar to that found by \citet{WD95} for the Orion nebula.  In the case of the spherical impostor, the true hydrogen density decreases away from the central star, while the apparent density deduced from the \dforba{O}{ii}{3726}{29} line ratio is constant within the first 70 pixels, because the emission lines are strongly weighted by the inner spherical skin of the O$^{+}$ zone.

The observed apparent electron temperature is remarkably similar in both models within the first 100 pixels. This however may be fortuitous, and might not occur for different density distributions, because the electron temperature depends on the relative proportions of different ions on the line of sight. In the outer parts, on the other hand, there is a conspicuous difference. While the apparent temperature continues to slowly decrease in the case of the blister, it significantly increases in the spherical impostor. This is due to a more important hardening of the ionizing radiation field in the latter case.

Finally, the derived N/O ratio is equal to the input value only in the outer part of the spherical impostor, where the flux is very small. In the brightest regions of both models, the apparent N/O ratio is smaller than the true one by significant amounts (about 0.25 and 0.65 times the true value in the spherical and blister models, respectively). The apparent N/O ratio for the entire nebulae is 0.68 (0.50) times smaller than the true one in the case of the blister (spherical impostor). 

Figures~\ref{fig:bli:hist} and~\ref{fig:sph:hist} show the 2D histogram
of the number of cells with a given value of $n_{\rm e}$~$n$(N$^+$) and $n_{\rm e}$~$n$(O$^+$)  within the nebula (blister and spherical impostor resp.), in the same way as in  Fig.~\ref{fig:res10}. In both cases, an important amount of the gas is in a situation where N$^+$/O$^+$=N/O (points on the first bisector), but a non negligible amount is also located far from this line, causing the discrepancy between the ``apparent'' and true values for N/O. Note that in the case of the impostor, the fact that the model is spherical leads to a very narrow line in the 2D histogram.

The main conclusion of this section is that the apparent properties of two models that differ only by their geometries and in such a way to conserve the   H$\beta$ surface brightness distribution may be significantly different. We also note that, in the cases we considered, the N/O ratio derived for the entire nebulae using the classical recipe that N/O=N$^+$/O$^+$ is significantly below the true value, even though N$^+$/N and O$^+$/O are not very small. In the present case, this is due to the spectral distribution of the ionizing radiation field (with realistic stellar atmospheres, the derived N/O would likely be different).
            \section{Conclusion and future work}
\label{sec:cl}

We have presented a quick pseudo-3D photoionization code, \nebu/,  and its associated visualization tool, VIS\_NEB3D.
This code is able to treat a large variety of problems in a time that is by orders of magnitude faster than the time required by proper 3D photoionization codes which treat the diffuse radiation in an exact manner.  \nebu/ is not able to correctly treat cases where the diffuse radiation field is important, such as shadows behind optically think clumps or very complex geometries. On the other hand, it is more accurate than 3D photoionization codes that work under the on-the-spot approximation for the diffuse ionizing radiation.

The main virtue of \nebu/ over accurate 3D photoionization codes is that it allows an efficient exploration of a large parameter space implied by non spherical geometries. This should be useful as a first approach for a tailored 3D photoionization modelling of real nebulae. Also, it allows one to easily run models with various geometries, and to get educated on the properties of such models which are not easy to predict without computations and therefore to exercise more care when interpreting observational data.

We have treated two examples to illustrate the possible applications of our code. One is the case of an ellipsoidal planetary nebula with two knots symmetrically located on the polar axis. We have shown and commented line intensity maps corresponding to different viewing angles. We have also used the computed line intensities to derive physical properties of the model in the same way as an observer would do for a real object. For example, we have derived an apparent value of N/O both for the entire nebula and along spectral slits of different orientations, assuming the common recipe that N/O=N$^+$/O$^+$. Interestingly, we found that, for this high excitation planetary nebula, for which one would expect the N/O to be less accurate than for lower excitation objects, the apparent N/O value is very close to the true one for all the observational set-ups.  

The other example is that of a blister HII region, and of its spherical impostor, i.e. a spherical model which has the same H$\beta$ flux surface brightness distribution, and the same ionizing star. We have shown that the properties of these two models differ drastically in many respects. We have also computed the apparent N/O ratios and found that they are significantly smaller (by factors 0.68 and 0.50 for the blister and the spherical impostor resp.) than the true values. 

These two examples should serve as a warning against preconceived ideas when interpreting spectroscopic and imaging data of planetary nebulae and HII regions.

The velocity fields of a blister HII region and of a spherical HII regions are however expected to be quite different.  They should allow one to discriminate various models. We will address this issue in a future paper. 

We plan to make the \nebu/ and VISNEB\_3D codes publicly available in due time.

\section*{Acknowledgments}

G.S. and M.P. benefited from a CNRS-CONACYT grant.
G.S. is grateful to the Instituto de Astronomia, UNAM which provided additional funds.
C.M. and G.S. are pleased to acknowledge  for hospitality the INAOE (Tonantzintla, Puebla, Mexico), where part of this work was carried out.
The computations are made on a AMD-64bit computer financed by grant PAPIIT IX125304 from DGAPA (UNAM, Mexico).

We thank the referee, Barbara Ercolano for her detailed comments.
%
%

\bibliography{uchii}

\bibliographystyle{mn2e}
\end{document}